\documentclass[preprint,12pt]{article}

\usepackage[font=small]{caption}
\usepackage{amssymb, amsmath, amsbsy} 
\usepackage{slashed}
\usepackage{graphicx}
\usepackage{float}
\usepackage{dsfont}
\usepackage{color}
\usepackage[affil-it]{authblk}
\usepackage{authblk}
\usepackage{rotating}
\usepackage{booktabs}
\usepackage[left=2cm,right=2cm,top=3cm,bottom=3cm]{geometry}
\usepackage{xcolor}

\newcommand{\g}{\gamma}
\newcommand{\s}{\slashed}
\newcommand{\la}{\langle}
\newcommand{\ra}{\rangle}

\author[1,2]{J.Lorenzo D\'iaz-Cruz\thanks{\texttt{jldiaz@fcfm.buap.mx}}}
\author[3,4]{Bryan O. Larios-L\'opez\thanks{\texttt{bryanlarios@gmail.com}}}
\title{\textbf{ Very long-lived Stop NLSP in MSSM scenarios with Gravitino LSP}}

\affil[1]{Centro Internacional de F\'isica Fundamental, BUAP
\protect\\Ciudad Universitaria, Puebla, Pue. M\'exico} 
\affil[2]{Facultad de Ciencias F\'isico - Matem\'aticas, BUAP \protect\\Apdo. Postal 1364, C.P. 72000, Puebla, Pue. M\'exico} 
\affil[3]{Departamento de Gravitaci\'on y Altas Energ\'ias 
\protect\\Facultad de Ciencias
\protect\\Universidad Nacional Aut\'onoma de Honduras
\protect\\Ciudad Universitaria, Tegucigalpa M.D.C. Honduras}
\affil[4]{Mesoamerican Centre for Theoretical Physics, Universidad Auto\'noma de Chiapas 
\protect\\Carretera Zapata Km. 4, Real del Bosque (Ter\'an), 29040, Tuxtla Guti\'errez, Chiapas, M\'exico}

\begin{document} 

\maketitle

\begin{abstract}
We present the calculation of the stop ($\tilde{t}$) lifetime that results from its decay   into gravitinos ($\tilde{\Psi}^{\mu}$) in the final state, namely $\tilde{t}\to \tilde{\Psi}^{\mu} + X$, where $X=t, bW, b l\nu_l$ for the 
two-, three- and four-body decay modes. The full calculation is obtained using the Feynman rules for massive gravitino, which is compared with the results obtained 
employing  the equivalence theorem, where the longitudinal component of the gravitino is replaced by the goldstino.
The stop turns out to be very long-lived in these scenarios, with lifetimes of $\mathcal{O}$($10^8$s, $10^{14}$s, $10^{20}$s) for  the corresponding 2-,3- and 4-body modes under consideration, and therefore all of them are safe from the big bang nucleosynthesis  problem. However, the lifetime for the four body decay mode occurs during the epoch of galaxy formation. When  the stop is produce at colliders, in these scenarios, it will hadronize and decay outside of the detector, even for the lowest values of stop lifetime.
\end{abstract}  


 \unitlength=1mm

\newpage

\section{Introduction}\label{introduction}

Supersymmetric (SUSY) theories have had a great impact 
in particle physics \cite{SUSYPrimer}.
Nevertheless, it remains an open issue how is  SUSY  realized in nature. 
The minimal supersymmetric extensions of the standard model (MSSM) is the simplest of such realization, it has been thoroughly studied and the search for its 
experimental signatures are the target of dedicated studies at the current LHC collider.
SUSY models with a discrete symmetry,  $R$ parity, 
assure the stability of the lightest supersymmetric particle (LSP) \cite{EHNOS}, 
in such case the LSP is a good candidate for dark matter 
(DM). Candidates for the LSP include sneutrinos, the
lightest neutralino $\chi^0_1$ \cite{vcmssm} and the gravitino $\tilde{\Psi}_{\mu}$ ~\cite{FOS94}.

In scenarios with gravitino LSP as DM candidate \cite{FengGDM,gdm,FengGDM2},  the nature of the next-to-lightest supersymmetric particle (NLSP) 
determines the relevant phenomenology~\cite{steffan, Bench3}.
Possible candidates for NLSP include the lightest neutralino~\cite{Neutralino,Neutralino2}, 
the chargino~\cite{chargino}, the lightest charged slepton~\cite{cslepton},  or  
the sneutrino~\cite{FengGDM3,KKKM,lcovi,lvelazco}.  Squark species could also play the role of NLSP, and in such case a natural candidates for NLSP 
could be the sbottom~\cite{ER,BDD,stopco} or the lightest stop $\tilde{t}_1$. Such  stop is found in most popular models of mediation for SUSY breaking, when the evolution of high scale  parameters are evolved down to lower energy scales. 

The NLSP has a long lifetime in these scenarios, due to the weakness of the gravitational interactions,  
and this leads to  a metastable charged sparticle that  could have dramatic signatures at
colliders~\cite{Are,Nojiri} and it could also affect  the big bang nucleosynthesis (BBN)~\cite{CEFO,KKM,kohri}.
There are several experimental and cosmological constraints for
the  scenarios with a gravitino LSP and a stop NLSP that were discussed in \cite{jlorenzo}. 
It turns out that the lifetime of the stop ${\tilde t_1}$ could be (very) long, in which case the
relevant collider limits are those on (apparently) stable charged particles. For instance the limits available from the 
Tevatron collider imply that $m_{\tilde t_1} > 220$~GeV~\cite{Tevatron}~\footnote[1]{The LHC will probably be sensitive to a
metastable ${\tilde t_1}$ that is an order of magnitude heavier.}.
Thus, knowing very precisely  the stop lifetime is one of the most important issues in this scenario, and this is 
precisely the goal of our work.

Depending on the masses of the stop ($m_{\tilde{t}_1}$) and gravitino ($\tilde{m}$), we should need to consider different modes. For $m_{\tilde{t}_1}>m_{t}+\tilde{m}$, it is the 2-body mode $\tilde{t}\to t\,\tilde{\Psi}_{\mu}$, while for $m_{t}+\tilde{m}>m_{\tilde{t}_1}>\tilde{m}+M_W+m_b$ we should consider the 3-body decay $\tilde{t}\to W\,b\tilde{\Psi}_{\mu}$. For $\tilde{m}+M_W+m_b>m_{\tilde{t}_{1}}>\tilde{m}+m_l+m_{\nu}$ the 4-body mode must be considered. It is expected that the stop lifetime will satisfy $\tau_{(2-body)}<\tau_{(3-body)}<\tau_{(4-body)}$, and the precise lifetime values will determine the possible role of these modes at colliders and cosmology. 
On one side, the stop lifetime will determine whether its decay could occur inside the detector when it is produced in a collider experiment, such as LHC. On the other side, the stop lifetime could reach values that may affect nucleosynthesis or the CMB, or in the extreme case it could even affect the early stages of galaxy formation.

Being the massive gravitino a spin-3/2 particle, the calculation of its decay modes or production reaction using the traditional methods with Feynman rules  present some difficulties. Thus, it   could be of great help to find alternative calculation method. This was initiated some time ago \cite{roy1,roy2,novaes}, but more recently we have resorted to the modern amplitude methods to simplify these calculations \cite{us_sugra}.

One of the starting points of modern approach to evaluate amplitudes using helicity methods is the known result that tree-level amplitudes including $n$ massless  gauge bosons of helicity configurations $(+,+,\dotsm,+)$ or $(-,-,\dotsm,-)$ vanish exactly; one needs to have at least two helicities of each sign in order to have a non-vanishing amplitude, i.e. $(-,-,+,+,\dotsm,+)$ or $(+,+,-,-,\dotsm,-)$. \\
On the other hand, processes involving massive gauge bosons are relevant to prove the mechanism of EWSB at LHC;  these include the production of massive particles, such as the W, Z gauge bosons, the heavy top quark and nowadays the Higgs boson. Thus, it would be interesting to extend the results of massless gauge boson scattering  to the massive case, and apply it to the massive gravitino process. \\

In this paper we present a detailed calculation of the stop lifetime, looking at each kinematical region where
two-, three- and four-body dominate.
Besides calculating the amplitude using the full wave function for the gravitino, we have also calculated the 
 decay width (and lifetime) using the
gravitino-goldstino equivalence theorem~\cite{goldstino}.

The organization of our paper goes as follows. After presenting these introductory ideas in Section \ref{introduction}, we present in Section  \ref{MSSM} some comments about the MSSM and the stop, moving then to discuss the gravitino wave function and the equivalence theorem. Then, in Section  \ref{stop_decays} we present the details of our calculations, starting with  a discussion of the relative size of the kinematical regions where the two, three- and four-body modes are allowed. We also present the amplitudes for each of these decay modes; the corresponding amplitudes using the equivalence theorem are presented too.
Finally, we present the numerical results for the stop lifetime  in Section \ref{numerical}, with our conclusions appearing in Section \ref{conclusions}. Some conventions and basics of amplitudes methods are containing for the Appendix.
  
\section{The MSSM, the stop and the Gravitino}\label{MSSM}
Supersymmetric quantum field theories are well appreciated in model building because of its improved UV behavior. Knowing that  SUSY-QFT are free of quadratic divergences, gave the hope that it could solve or at least ameliorate the hierarchy/naturalness problem.
However, to construct realistic models one must impose that SUSY is broken, which introduces some complications and problems. Depending on the mechanism of mediation of SUSY breaking, one can have different models with their own phenomenological consequence.  
\subsection{The MSSM and the stop squark}
The MSSM minimal implementation of SUSY in the SM, which has the same gauge group and matter content as in the SM, but this time the different fields are replaced by the corresponding superfields. Thus, gauge bosons have gauginos as superpartners, fermions come with sfermions and Higgs with the higgsinos.  

A consistent formulation of local SUSY leads to Supergravity (SUGRA), where graviton (spin-2) comes with its superpartner, the gravitino with spin-3/2. SUGRA Lagrangian includes interactions of the type: $\tilde{f}\Psi_{\mu}A^{\mu},\,f\tilde{f}\Psi_{\mu},\,\tilde{f} f\Psi_{\mu}A^{\mu}$, for fermion ($f$), sfermion ($\tilde{f}$), gauge boson ($A^{\mu}$) and gravitino ($\Psi_{\mu}$). To calculate processes involving the gravitino, we need to write such Lagrangian in terms of mass-eigenstates, which is covered in the literature \cite{moroi}.

We present  some  relevant formulae for the input parameters that appear in the Feynman rules of the gravitino
within the MSSM.
The (2x2) stop mass matrix can be written as:

\begin{equation}
\widetilde{M}_{\tilde{t}}^2 =
 \begin{pmatrix}
  M_{LL}^2  & M_{LR}^2 \\
 M_{LR}^{2\,\dag} & M_{RR}^2\\
 \end{pmatrix},
 \label{eq:01}
 \end{equation}

where the entries take the form:

\begin{align} \nonumber
M_{LL}^2 &= M_{L}^2+m_t^2+\frac{1}{6}\cos2\beta \,(4M_W^2-m_Z^2),\\
M_{RR}^2 &= M_{R}^2+m_t^2+\frac{2}{3}\cos2\beta\sin^2\theta_W\, m_Z^2,\\ \nonumber
M_{LR}^2 &= -m_t (A_t + \mu \, \cot \beta) \equiv - m_t X_t\,.
\end{align}

The corresponding mass eigenvalues are given by:

\begin{equation}
m^2_{\tilde{t}_1}=m_t^2 + \frac{1}{2}(M_{L}^2+  M_{R}^2)+
\frac{1}{4}m^2_Z \cos 2\beta-\frac{\Delta}{2}   ,   
\end{equation}

and

\begin{equation}
m^2_{\tilde{t}_2}= m^2_t + \frac{1}{2}(M_{L}^2+  M_{R}^2)+
\frac{1}{4}m^2_Z \cos 2\beta+\frac{\Delta}{2}    ,            
\end{equation}

where $\Delta^2= \left( M_{L}^2 -  M_{R}^2 + \frac{1}{6} \cos 2\beta (8
m^2_W-5m^2_Z) \right)^2 + 4\, m_t^2 |A_t + \mu \cot \beta |^2$.
The mixing angle $\theta_{\tilde{t}}$ appears in the   mixing matrix that relate the 
weak basis $(\tilde{t}_L,\tilde{t}_R)$ and the mass
eigenstates  $(\tilde{t}_1,\tilde{t}_2)$, and it is given by
$\tan \theta_{\tilde{t}}= \frac{(m^2_{\tilde{t}_1}-M^2_{LL})}{|M^2_{LR}|}$.
From these expressions one can see that
in order to obtain a very light stop one needs to have a
very large value for the term $M^?{LR}$, which can happen when the trilinear soft supersymmetry-breaking parameter is large~\cite{stopco,ben}. 
It turns out that  such scenario helps to produce a Higgs mass value in agreement with the mass 
measured at LHC (125-126 GeV), in a consistent way within the MSSM. 

\subsection{Gravitino wave functions}\label{gravitino_theory}

The Rarita-Schwinger equation that describes a massive spin-3/2 particle \cite{moroi,rarita,auvil} includes the following set of equations

\begin{align}
\g_{\mu}\tilde{\Psi}^{\mu}_{\lambda_p}(p)&=0,\label{eq1:irreduciblecondition}\\
p_{\mu}\tilde{\Psi}^{\mu}_{\lambda_p}(p)&=0,\label{eq2:irreduciblecondition}\\
(\s{p}-\tilde{m})\tilde{\Psi}^{\mu}_{\lambda_p}(p)&=0.\label{eq3:irreduciblecondition}
\end{align}

We shall start by writing the wave function for the four polarization states ($\lambda_p=\pm\frac{3}{2},\,\pm\frac{1}{2}$) of the gravitino ($\tilde{\Psi}_{\lambda_p}^{\mu}(p)$) in momentum space which fulfill 
these equations, in terms of spin-1 and spin-1/2 components as follows

\begin{align}
\tilde{\Psi}_{++}^{\mu}(p)&=\epsilon_{+}^{\mu}(p)u_+(p),\label{eq:gravitinostate01}\\
\tilde{\Psi}_{--}^{\mu}(p)&=\epsilon_{-}^{\mu}(p)u_-(p),\label{eq:gravitinostate02}\\
\tilde{\Psi}_{+}^{\mu}(p)&=\sqrt{\frac{2}{3}}\epsilon_{0}^{\mu}(p)u_+(p)+\frac{1}{\sqrt{3}}\epsilon_{+}^{\mu}(p)u_-(p),\label{eq:gravitinostate03}\\
\tilde{\Psi}_{-}^{\mu}(p)&=\sqrt{\frac{2}{3}}\epsilon_{0}^{\mu}(p)u_-(p)+\frac{1}{\sqrt{3}}\epsilon_{-}^{\mu}(p)u_+(p)
\label{eq:gravitinostate04},
\end{align}

where $\epsilon_{\pm}^{\mu}(p)$ are the transversal d.o.f. of the gravitino, i.e. the polarization vector for the corresponding helicities $\lambda=\pm$, furthermore, $\epsilon_{0}^{\mu}(p)$ is the longitudinal d.o.f corresponding to the helicity $\lambda=0$. The $u_{\pm}(p)$ are the massive Dirac spinor for the labels $\lambda=\pm$, which would correspond to the helicities in massless case.

\subsection{Spinor Helicity Formalism for massive spin-$3/2$ gravitino field}\label{helicitys_pinor_formalism}
In order to compute Scattering Amplitudes (SA)  with  massive spin-3/2 gravitino field in the final state, 
we shall make use the Spinor Helicity Formalism (SHF) \cite{schwartz, srednicki, elvang, bryan,lorenzo} which has great advances in order to handle perturbative calculations in quantum field theories. In this paper we want to compute SA for massive gravitino. Here, we use  the Light Cone Decomposition (LCD) technique \cite{boels, weinzierl, spinorsextras}, in order to express massive momenta in terms of massless ones. In the Appendix  we review   some basic properties of the massless SHF that will also be useful as a starting point for the massive case.  

In modern SHF, the polarization vectors $\epsilon_{\pm}^{\mu}(p),\,\epsilon_{0}^{\mu}(p)$  as well as the massive Dirac spinors $u_{\pm}(p)$ are written in terms of  bra-ket notation \cite{dittmaier}.
Then, it is straightforward to express  the four gravitino states in this bra-ket notation, they are given as follows:  

\begin{align}
\tilde{\Psi}^{\mu}_{++}(p)&=\frac{\la r|\g^{\mu}|q]}{\sqrt{2}[rq]}\left(|r\ra+\tilde{m}\frac{|q]}{[rq]}\right),\\
\tilde{\Psi}^{\mu}_{--}(p)&=\frac{\la q|\g^{\mu}|r]}{\sqrt{2}\la rq\ra}\left(|r]+\tilde{m}\frac{|q\ra}{\la rq\ra}\right),\\
\tilde{\Psi}^{\mu}_{-}(p)&=\sqrt{\frac{2}{3}}\left(\frac{r^{\mu}}{\tilde{m}}-\tilde{m}\frac{q^{\mu}}{s_{qr}}\right)\left(|r]+\tilde{m}\frac{|q\ra}{\la rq\ra}\right)+\frac{1}{\sqrt{3}}\frac{\la q|\g^{\mu}|r]}{\sqrt{2}\la rq\ra}\left(|r\ra+\tilde{m}\frac{|q]}{[rq]}\right),\\
\tilde{\Psi}^{\mu}_{+}(p)&=\sqrt{\frac{2}{3}}\left(\frac{r^{\mu}}{\tilde{m}}-\tilde{m}\frac{q^{\mu}}{s_{qr}}\right)\left(|r\ra+\tilde{m}\frac{|q]}{[rq]}\right)+\frac{1}{\sqrt{3}}\frac{\la r|\g^{\mu}|q]}{\sqrt{2}[rq]}\left(|r]+\tilde{m}\frac{|q\ra}{\la rq\ra}\right),
\end{align}

where the momentum $p$ is defined as $p=r-\frac{\tilde{m }}{2r\cdot q}q$, with the momenta $r^{\mu}$ and $q^{\mu}$  being massless and  $|r\ra$ and  $|q\ra$ denoting their associated two component momentum spinor. The Mandelstam-like variable $s_{qr}$ is  defined as $s_{qr}=-(q+r)^2=-2q\cdot r$. For our goals, it shall be useful to rearrange  the four gravitino states as an expansion in terms of the gravitino mass $(\tilde{m})$, namely:  

\begin{align}
\tilde{\Psi}_{++}^{\mu}(p)&=\beta_1^{\mu}|r\rangle+\tilde{m}\beta_2^{\mu}|q],\label{eq:gravitinobasis1}\\
\tilde{\Psi}_{--}^{\mu}(p)&=-\beta_1^{*\mu}|r]+\tilde{m}\beta_2^{*\mu}|q\rangle,\\
\tilde{\Psi}_-^{\mu}(p)&=\frac{1}{\tilde{m}}\beta_3^{\mu}|r]+\beta_4^{\mu}|q\rangle+\beta_5^{\mu}|r\rangle+\tilde{m}(\beta_6^{\mu}|r]+\beta_7^{\mu}|q])+\tilde{m}^2\beta_8^{\mu}|q\rangle,\\
\tilde{\Psi}_+^{\mu}(p)&=\frac{1}{\tilde{m}}\beta_3^{*\mu}|r\ra-(\beta_4^{*\mu}|q]+\beta_5^{*\mu}|r])+\tilde{m}(\beta_6^{*\mu}|r\ra+\beta_7^{*\mu}|q\ra)-\tilde{m}^2\beta_8^{*\mu}|q].
\label{eq:gravitinobasis4}
\end{align}

The gravitino mass $\tilde{m}$ is directly connected with the the SUSY breaking energy scale $F$ as $\tilde{m}=\frac{F}{\sqrt{3}M}$, where $M$ denotes the Plank mass. 
The expressions for all the $\beta_i^{\mu}\,(\forall\,i=1\ldots8)$ are shown in  Table \ref{betas}.

\begin{table}[H]
\begin{center}
  \begin{tabular}{  || c | c |c || }
    \hline \hline
   $i$  &  $\beta_i^{\mu}$ & $\beta_i^{*\mu}$ \\ \hline
   \hline\hline
   \large
    $1$ &\large $\frac{\la qr\ra \la r |\g^{\mu}|q]}{\sqrt{2}s_{qr}}$ & $\frac{ [rq] \la q |\g^{\mu}|r]}{\sqrt{2}s_{qr}}$ \\ \hline 
    2 &  $\frac{\la qr\ra^2 \la r |\g^{\mu}|q]}{\sqrt{2}s_{qr}^2}$ &  $\frac{ [rq]^2 \la q |\g^{\mu}|r]}{\sqrt{2}s_{qr}^2}$ \\ \hline
     3 &  $ \sqrt{\frac{2}{3}}r^{\mu}$ &  $\sqrt{\frac{2}{3}} r^{\mu}$ \\ \hline
      4 &  $\sqrt{\frac{2}{3}}\frac{[qr]r^{\mu}}{s_{qr}}$ &  $\sqrt{\frac{2}{3}}\frac{ \la rq\ra r^{\mu}}{s_{qr}}$\\ \hline
       5 &  $\sqrt{\frac{2}{3}}\frac{[qr]\la q| \g^{\mu}| r]}{2s_{qr}}$ &  $\sqrt{\frac{2}{3}}\frac{ \la rq\ra \la r| \g^{\mu}| q]}{2s_{qr}}$  \\ \hline
        6 &  $-\sqrt{\frac{2}{3}}\frac{ q^{\mu}}{s_{qr}}$ &  $-\sqrt{\frac{2}{3}}\frac{ q^{\mu}}{s_{qr}}$ \\ \hline
         7 &  $-\sqrt{\frac{2}{3}}\frac{\la q |\g^{\mu}|r]}{2s_{qr}}$ &  $-\sqrt{\frac{2}{3}}\frac{ \la r |\g^{\mu}|q]}{2s_{qr}}$ \\ \hline
          8 &  $-\sqrt{\frac{2}{3}}\frac{ q^{\mu}[qr]}{s_{qr}^2}$ &   $-\sqrt{\frac{2}{3}}\frac{ q^{\mu}\la rq\ra}{s_{qr}^2}$ \\ \hline
\hline
  \end{tabular}
    \caption{Definitions of the $\beta_{i}^{\mu}$ $\forall\,i=1\ldots8$ with $s_{qr}=-(q+r)^2$.} 
  \label{betas}
 \end{center}
\end{table}

Just for completeness we also show the wave functions corresponding to the four gravitino conjugate states $\overline{\tilde{\Psi}}^{\mu}_{\lambda_p}(p)$ with $\lambda_p=++,--,+,-$, which take the following form:

\begin{align}
\overline{\tilde{\Psi}}_{++}^{\mu}(p)&=\beta_1^{*\mu}[r|+\tilde{m}\beta_2^{*\mu}\la q|,\label{eq:gravitinobasis5}\\
\overline{\tilde{\Psi}}_{--}^{\mu}(p)&=-\beta_1^{\mu}\la r|+\tilde{m}\beta_2^{\mu}[q|,\\
\overline{\tilde{\Psi}}_-^{\mu}(p)&=\frac{1}{\tilde{m}}\beta_3^{*\mu}\la r|+\beta_4^{*\mu}[q|+\beta_5^{*\mu}[r|+\tilde{m}(\beta_6^{*\mu}\la r|+\beta_7^{*\mu}\la q|)+\tilde{m}^2\beta_8^{*\mu}[q|,\\
\overline{\tilde{\Psi}}_+^{\mu}(p)&=\frac{1}{\tilde{m}}\beta_3^{\mu}[r|-(\beta_4^{\mu}\langle q|+\beta_5^{\mu}\langle r|)+\tilde{m}(\beta_6^{\mu}[r|+\beta_7^{\mu}[q|)-\tilde{m}^2\beta_8^{\mu}\langle q|.\label{eq:gravitinobasis8}
\end{align}

Having expressed the massive gravitino states in this  basis makes even  simpler to handle the helicity amplitudes. We can check that if the four gravitino states in this new notation fulfill the equations (\ref{eq1:irreduciblecondition})-(\ref{eq3:irreduciblecondition}),  as well as the normalization condition

\begin{equation}\label{normalization_gravitinos}
\overline{\tilde{\Psi}}_{\lambda_1\mu}\tilde{\Psi}_{\lambda_2}^{\mu}=2\tilde{m}\lambda_{\lambda_1\lambda_2}.
\end{equation}

 For example, we can verify that the gravitino states fulfill the normalization condition Eq.~(\ref{normalization_gravitinos}),  for $\lambda_1=\lambda_2=-$, namely:

\begin{align}
\overline{\tilde{\Psi}}^{\mu}_-(p)\tilde{\Psi}_{\mu-}(p)&=\la rq\ra\big(\beta_3^{*\mu}\beta_{4\mu}+\beta^{*\mu}_{3}\beta_{8\mu}\tilde{m}^3+\beta_{6}^{*\mu}\beta_{4\mu}\tilde{m}^3+\beta_{6}^{*\mu}\beta_{8\mu}\tilde{m}^5-\beta_{7}^{*\mu}\beta_{5\mu}\tilde{m}^3\big)+\text{c.c.}\\
&=\la rq\ra\Big(-\frac{4[qr](r\cdot q)}{3s_{rq}^2}\tilde{m}^3-\frac{[qr]}{3s_{qr}}\tilde{m}^3\Big)+\text{c.c.}\\
&=2\tilde{m}.
\end{align}

As  can be seen from the last step, the equations (\ref{eq:gravitinobasis1})-(\ref{eq:gravitinobasis4}) and (\ref{eq:gravitinobasis5})-(\ref{eq:gravitinobasis8}) are very convenient in order to handle the  messy algebraic expressions that appear in  processes  involving the massive gravitino.

\subsection{The light gravitino and equivalence theorem}\label{goldstino}
According  to the equivalence theorem, in the high-energy limit  on-shell scattering amplitudes involving longitudinal vector bosons can be calculated by replacing 
them by the corresponding goldstone bosons. Thus, the amplitude summed over polarizations ($W=W_{L},W_{T}$) is

\begin{equation}
 \sum_{pol}\mathcal{A}_i( W_i(p_i) \cdots )=\mathcal{A}_{0}(W_{T}(p_i)\cdots)+\mathcal{A}_{1}(W_{L}(p_i)\cdots).
\end{equation}

Then, the equivalence theorem allows us to write:
$ \mathcal{A}_{1}(W_L(p_i) \cdots )=\mathcal{A}_{1}( \phi_W(p_i) \cdots )$, where $\phi_W(p_i)$ denotes the goldstone boson that replaces the longitudinal component  of the massive gauge boson.

A similar result holds for the massive gravitino scattering \cite{goldstino}. Namely, some amplitudes with certain helicity configurations that vanish in the massless case, would get corrections of  order $\mathcal{O}(\frac{\tilde{m}}{E})$, where $\tilde{m}$ denotes the gravitino mass, and $E$ is the typical energy of the physical process.  Again, this can be evaluated by relying on the SUGRA equivalence theorem.  We are working with the $\pm$ helicity states associated with the goldstino that arise from the Super-Higgs mechanism, which is required to break Supersymmetry and to induce masses for the superpartners including the gravitino.\\

For the strict massless case, one can simply apply the massless helicity methods,
while for the massive case, one requires to take into account the massive Dirac equation and the light-cone decomposition.
We consider the gravitino 4-momentum in spherical coordinates 

\begin{equation}\label{eq:gmomentum}
p^{\mu}=(E,|\vec{p}|\sin\theta\cos\phi,|\vec{p}|\sin\theta\sin\phi,|\vec{p}|\cos\theta) ,
\end{equation}

with $p^2=-\tilde{m}^2$. The polarization vectors take the following form

\begin{align}
\epsilon_{+}^{\mu}(p)&=\frac{1}{\sqrt{2}}(0,\cos\theta\cos\phi-i\sin\phi,\cos\theta\sin\phi+i\cos\phi,-\sin\theta),\\
\epsilon_{-}^{\mu}(p)&=-\frac{1}{\sqrt{2}}(0,\cos\theta\cos\phi+i\sin\phi,\cos\theta\sin\phi-i\cos\phi,-\sin\theta),\\
\epsilon_{0}^{\mu}(p)&=-\frac{1}{\tilde{m}}(|\vec{p}|,-E\sin\theta\cos\phi,-E\sin\theta\sin\phi,-E\cos\theta).\label{eq:polarization_zero}
\end{align}

In the limit $|\vec{p}|\to\infty$, one has that $E\approx|\vec{p}|$, which implies that

\begin{align}
\epsilon_{\pm}^{\mu}(p)p_{\mu}&=-\epsilon_{\pm}^0(p)p_0+\vec{\epsilon}_{\pm}(p)\cdot\vec{p}\\
&=-\epsilon_{\pm}^{0}(p)|\vec{p}|+|\vec{\epsilon}_{\pm}(p)||\vec{p}|\cos\theta.
\end{align}

 Using the transversality condition  $\epsilon^{\mu}_{\pm}(p)p_{\mu}=0$ in the above expression implies  $\epsilon_{+}^{\mu}(p)\approx0$ and  $\epsilon_{-}^{\mu}(p)\approx0$ when $|\vec{p}|\to\infty$. However,  in this limit the polarization vector $\epsilon_{0}^{\mu}(p)$ (\ref{eq:polarization_zero}) has the following expression:
 
 \begin{equation}
\epsilon_{0}^{\mu}(p)=\frac{p^{\mu}}{\tilde{m}}.
\end{equation}

Thus, the helicity states of the gravitino Eqs.~(\ref{eq:gravitinostate01})-(\ref{eq:gravitinostate04}) are reduced in the high energy limit, namely 

\begin{align}\label{eq:gfappox1}
\tilde{\Psi}_{++}^{\mu}(p)&\approx\mathcal{O}\left(\frac{\tilde{m}}{E}\right)\approx0,\\\label{eq:gfappox2}
\tilde{\Psi}_{--}^{\mu}(p)&\approx\mathcal{O}\left(\frac{\tilde{m}}{E}\right)\approx0,\\\label{eq:gfappox3}
\tilde{\Psi}_{-}^{\mu}(p)&\approx\sqrt{\frac{2}{3}}\epsilon_{0}^{\mu}(p)u_-(p)+\mathcal{O}\left(\frac{\tilde{m}}{E}\right)\approx\sqrt{\frac{2}{3}}\left(\frac{p^{\mu}}{\tilde{m}}\right)u_-(p),\\\label{eq:gfappox4}
\tilde{\Psi}_{+}^{\mu}(p)&\approx\sqrt{\frac{2}{3}}\epsilon_{0}^{\mu}(p)u_+(p)+\mathcal{O}\left(\frac{\tilde{m}}{E}\right)\approx\sqrt{\frac{2}{3}}\left(\frac{p^{\mu}}{\tilde{m}}\right)u_+(p),
\end{align}

such that the surviving  gravitino states are only those of helicity $\pm1/2$. For the transformation of Eqs.~(\ref{eq:gfappox3})-(\ref{eq:gfappox4})  into  coordinate space we need to replace $p^{\mu}\to i\partial^{\mu}$ in the gravitino field Eqs.~(\ref{eq:gfappox3})-(\ref{eq:gfappox4}), i.e. $\tilde{\Psi}_{\mu}(x)\to i\sqrt{\frac{2}{3}}\frac{\partial_{\mu}{\psi(x)}}{\tilde{m}}$, where $\psi(x)$ is the so-called spin-1/2 goldstino  state. After replacing the  gravitino field as goldstino approximation in the  Lagrangian with gravitino  $\Psi^{\mu}(x)$,  one obtains an effective Lagrangian describing the interaction of the goldstino with chiral superfields, which  is given by \cite{moroi}:

\begin{equation}\label{eq:goldstinolagrangian}
 \mathcal{L}=\frac{i(m_{\phi}^2-m_{\chi}^2)}{\sqrt{3}\tilde{m}M}(\bar{\psi}\chi_R)\phi^*-\frac{im_{\lambda}}{8\sqrt{6}\tilde{m}M}\bar{\psi}[\gamma^{\mu},\gamma^{\nu}]\lambda^{(a)}F^{(a)}_{\mu\nu}+h.c.
\end{equation}

In this approximation, one assembles the HAs from the Feynman rules considering the goldstino field as a Dirac spinor for computational purposes.

\bigskip

\section{Two-, three- and four-body stop decays }\label{stop_decays}

The stop decay modes  that include a gravitino in the final state, could include 2-body, 3-body or 4-body final states, which are:  $\tilde{t}\to t\tilde{\Psi}^{\mu}$, $\tilde{t}\to \tilde{\Psi}^{\mu}bW$ and $\tilde{t}\to \tilde{\Psi}^{\mu}b\nu \bar{l}$ or $\tilde{t}\to \tilde{\Psi}^{\mu}b q \bar{q}$.
The two-body mode is allowed in the kinematical regions $m_{\tilde{t}} \geq \tilde{m} + m_t$, while the three-body mode
becomes relevant in the range: $\tilde{m} + m_t \geq m_{\tilde{t}} \geq \tilde{m} + m_b + m_W $, and finally one needs to
consider the 4-body mode in the range: $ \tilde{m} + m_b + m_W \geq m_{\tilde{t}} \geq \tilde{m} +m_b$ 
(neglecting leptons masses). The relative sizes of these regions are shown in Figure \ref{parameter_space234} in the plane $\tilde{m}$-$m_{\tilde{t}}$.

\vspace{0.01\linewidth}

\begin{minipage}{\linewidth}
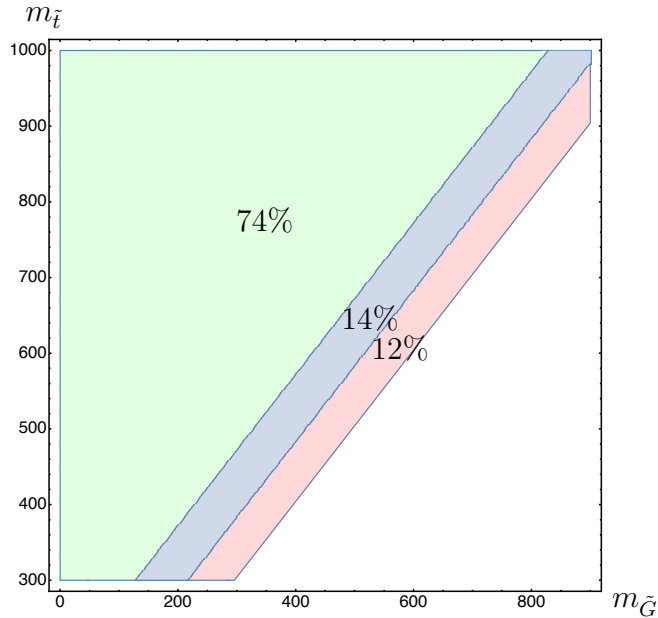
\begin{figure}[H]
\centering
\begin{picture}(-40,75)
\put(-70,0){\includegraphics[scale=0.4]{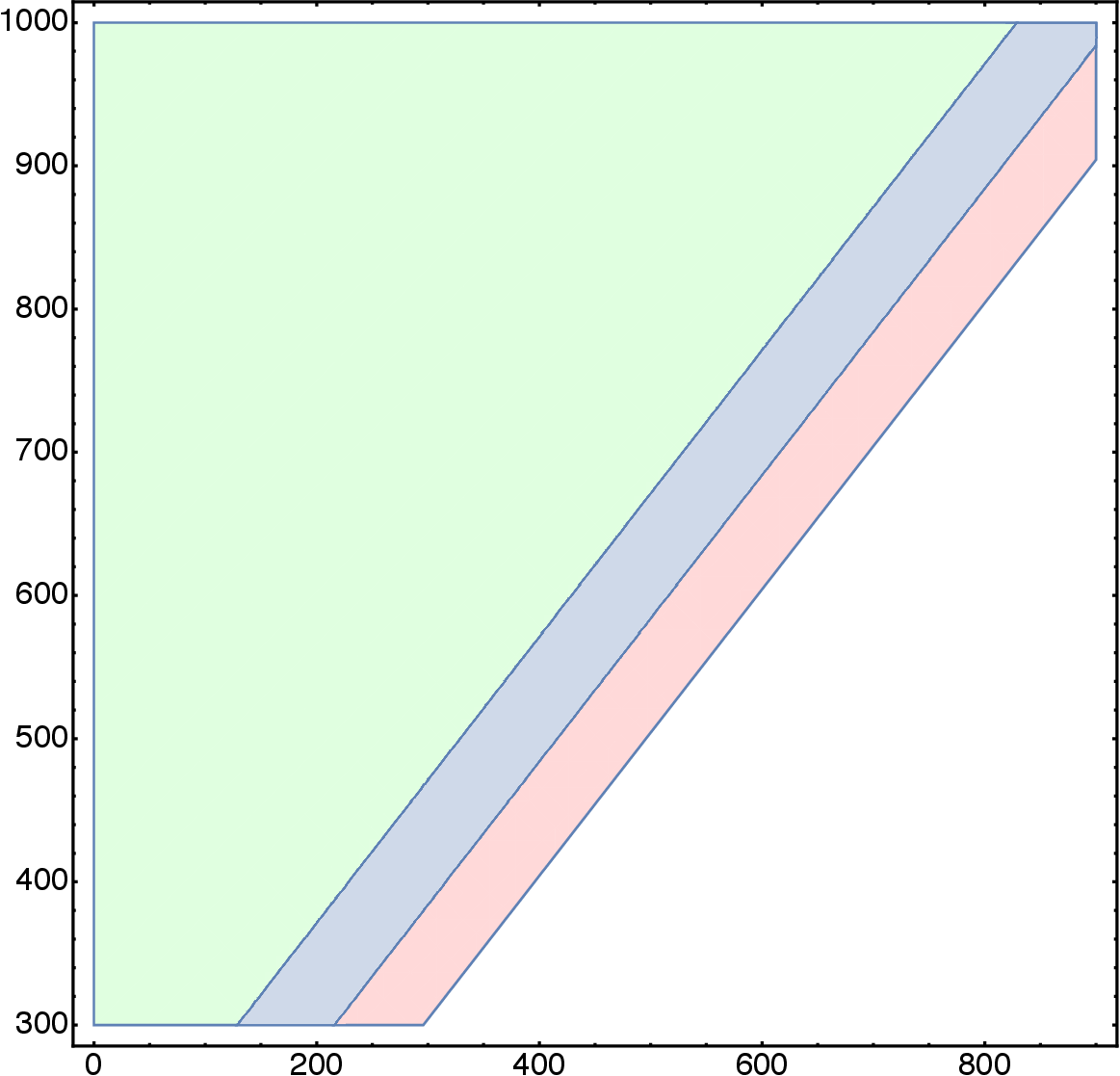}}
\put(-68,78){$m_{\tilde{t}}$}
\put(10,0){$m_{\tilde{G}}$}
\put(-22,33){$12\%$}
\put(-26,37){$14\%$}
\put(-40,50){$74\%$}
\end{picture}
\caption{Parameter Space for the two-,three and four-body decays.}
        \label{parameter_space234}
\end{figure}
      \end{minipage}
     
       \vspace{0.03\linewidth}

 As we can see from Figure \ref{parameter_space234}, the kinematical region (phase-space) in the plane $m_{\tilde{t}}-\tilde{m}$ for the two-body mode $\tilde{t}\to t\tilde{\Psi}^{\mu}$ is the largest one ($74\%$), but the ones where the three-body $\tilde{t}\to b W \tilde{\Psi}^{\mu}$ ($14\%$) and four-body modes $\tilde{t}\to b \bar{l}\nu_{l} \tilde{\Psi}^{\mu}$ occur (12\%) are not negligible at all. Thus, it is certainly relevant to determine the stop lifetime in each of these regions. 

Here, we shall study the stop decay in each region separately. However, it should be mentioned that by a proper treatment of the Breit-Wigner resonances it should be possible to have one simple formulae valid for all the kinematical regions. We present each case in order to compare our results with the literature, when available, and also to illustrate the  power of amplitude methods to treat the multi-particle states, in increasing order of complexity. 

\subsection{Full expressions for two-body stop decay with 
LSP gravitino  in the final state}

In this section we review the calculation of the two body stop decay  $\tilde{t}(p_1)\to \tilde{\Psi}^{\mu}(p_2)t(p_3)$ for the massive gravitino, and compare with the result of equivalence theorem. The amplitude for this decay takes the following form:

\begin{align}
\mathcal{M}=C_2\mathcal{T}_{\lambda_2\lambda_3},
\end{align}

the function $\mathcal{T}^{F}_{\lambda_2\lambda_3}$ for the full massive gravitino is as follows

\begin{equation}
\mathcal{T}^{F}_{\lambda_2\lambda_3}=\overline{\tilde{\Psi}}^{\mu}_{\lambda_2}(p_2)\g_{\alpha}\g_{\mu}p_1^{\alpha}(\mathbf{P_{R}}\cos\theta+\mathbf{P_{L}}\sin\theta)u_{\lambda_3}(p_3),
\end{equation}

with $C_2^{F}=\frac{g_W}{\sqrt{3}M}$, $g_W$ is the electroweak coupling constant and $M=M_{Pl}/\sqrt{8\pi}$ is the reduced Plank mass, with $M_{Pl}=1.2\times10^9$ GeV.  We are using $\mathbf{P_R}$ and $\mathbf{P_L}$ as the right and left projectors.
The  labels $\lambda_2\,(=-,+,--,++)$ and $\lambda_3\,(=-,+)$ denote the helicity labels for gravitino and top particles. According to the combinatoric of the helicity labels, there are eight HAs, but it can be shown that six of them vanish; thus only two nonzero HAs are left, which are shown in Table \ref{table3}.

\begin{table}[H]
\begin{center}
  \begin{tabular}{  || c |  c | c || }
    \hline \hline
    $\lambda_2\lambda_3$ &$\mathcal{T}_{\lambda_2\lambda_{3}}^{F}$ & $\mathcal{T}_{\lambda_2\lambda_{3}}^{E}$ \\ \hline
   \hline\hline
     $-,+$ & $\left(\frac{s_{r_2q_2}^2-\tilde{m}^2m_t^2}{s_{r_2q_2}\tilde{m}[r_2q_2]}\right)F_1$ & $\frac{1}{[q_2r_2]}F_1$\\ \hline
     $+,-$ & $\left(\frac{s_{r_2q_2}^2-\tilde{m}^2m_t^2}{s_{r_2q_2}\tilde{m}\langle r_2q_2\rangle}\right)F_2$ & $\frac{1}{\langle q_2r_2\rangle}F_2$ \\ \hline
    \hline
  \end{tabular}
  \caption{Helicity Amplitudes functions for the two-body stop decay ($\tilde{t}(p_1)\to \tilde{\Psi}^{\mu}(p_2)t(p_3)$) with LSP gravitino in the final state ($\mathcal{T}_{\lambda_2\lambda_{3}}^{F}$). We have also included the corresponding helicity amplitude  functions obtained using the equivalence theorem ($\mathcal{T}_{\lambda_2\lambda_{3}}^{E}$).}
  \label{table3}
 \end{center}
\end{table}

The corresponding function for the goldstino approximation ($\mathcal{T}^{E}_{\lambda_2\lambda_3}$) is given by

\begin{equation}
\mathcal{T}^{E}_{\lambda_2\lambda_3}=\bar{u}(p_2)(\mathbf{P_{R}}\cos\theta+\mathbf{P_{L}}\sin\theta)u(p_3),
\end{equation}
the coefficient appearing in the amplitude ($\mathcal{M}=C_2\mathcal{T}_{\lambda_2\lambda_3}$) is given by
with $C_2^{E}=\frac{g_W(m_{\tilde{t}}^2-m_t^2)}{\sqrt{3}M\tilde{m}}$. In Table \ref{table3}, we have defined the following functions
\begin{align}
F_1&=m_t\tilde{m}\sin\theta_{\tilde{t}} + s_{r_2q_2}\cos\theta_{\tilde{t}},\\
F_2&=m_t\tilde{m}\cos\theta_{\tilde{t}} + s_{r_2q_2}\sin\theta_{\tilde{t}}.
\end{align}

Here, $s_{r_2q_2}=-(r_2+q_2)^2$ is a Mandelstam-like variable; the $\theta_{\tilde{t}}$ denotes the mixing angle in the stop system.
The squared and averaged amplitude of the process $\tilde{t}(p_1)\to \tilde{\Psi}^{\mu}(p_2)t(p_3)$ corresponds to the sum of  the squared HA's shown in Table \ref{table3}, this reads as:

\begin{align}
\langle |\mathcal{M}^{F}|^2 \rangle&=|\mathcal{M}^F_{-,+}|^2+|\mathcal{M}^F_{+,-}|^2\\
&=\frac{(\tilde{m}^2m_t^2-s_{r_2q_2}^2)^2}{3M^2\tilde{m}^2s_{r_2q_2}^3}(m_t^2\tilde{m}^2+s_{r_2q_2}^2+2\sin2\theta_{\tilde{t}}\,m_t\tilde{m}s_{r_2q_2}).
\end{align} 

The HA's corresponding to the goldstino aproximaition are shown in the third column of Table \ref{table3}.
In this case the squared and averaged amplitude has the following form

\begin{align}
\langle |\mathcal{M}^E|^2 \rangle&=|\mathcal{M}^E_{-,+}|^2+|\mathcal{M}^E_{+,-}|^2\\
&=\frac{(m_t^2-m_{\tilde{t}}^2)^2}{3M^2\tilde{m}^2s_{r_2q_2}}(m_t^2\tilde{m}^2+s_{r_2q_2}^2+2\sin2\theta_{\tilde{t}} \,m_t\tilde{m}s_{r_2q_2}).
\end{align} 

\subsection{Three-body stop decay: $\tilde{t}\to\tilde{\Psi}^{\mu}Wb$}

Here we shall consider only the dominant contribution to the amplitude, with the top as intermediate state (Figure \ref{fig:tbsd}); this approximation should work very well for split/slim SUSY scenarios \cite{slim1,slim2}, where most scalar superpartners are much heavier than the stop and gravitino particles.
We have written the amplitude  for the three-body stop decay $\tilde{t}(p_1)\to\tilde{\Psi}^{\mu}(p_2)b(p_3)W(p_4)$ as follows

\begin{align}
\mathcal{M}_{\lambda_2\lambda_3\lambda_4}&=C_3P_{t}(l)\mathcal{T}_{\lambda_2\lambda_3\lambda_4},
\end{align}

the function $\mathcal{T}_{\lambda_2\lambda_3\lambda_4}^{F}$ for the full gravitino wave function takes the following form

\begin{align}
\mathcal{T}_{\lambda_2\lambda_3\lambda_4}^{F}&=\left(\overline{\tilde{\Psi}}^{\mu}_{\lambda_2}(p_2)\g_{\alpha}\g_{\mu}p_{1}^{\alpha}(\mathbf{P_{R}}\cos\theta+\mathbf{P_{L}}\sin\theta)\right)\left(-\s{l}+m_{t}\right)\left(\g_{\nu}\mathbf{P_{R}}u_{\lambda_3}(p_3)\epsilon^{\nu}_{\lambda_4}(p_4)\right),
\end{align}

with  $C_3^{F}=\frac{g_W}{2M}$. The denominator of the quark top propagator (Figure \ref{fig:tbsd}) is defined as $P_t(l)=\frac{1}{l^2+m_t^2}$. 
For the goldstino approximation the function $\mathcal{T}_{\lambda_2\lambda_3\lambda_4}^{E}$ is given as follows

\begin{align}
\mathcal{T}_{\lambda_2\lambda_3\lambda_4}^{E}&=\left(\bar{u}_{\lambda_2}(p_2)(\mathbf{P_{R}}\cos\theta+\mathbf{P_{L}}\sin\theta)\right)\left(-\s{l}+m_{t}\right)\left(\g_{\nu}\mathbf{P_{R}}u_{\lambda_3}(p_3)\epsilon^{\nu}_{\lambda_4}(p_4)\right),
\end{align}

with $C_3^{E}=\frac{g_W(m_t^2-m_{\tilde{t}}^2)}{2\sqrt{6}M\tilde{m}}$.

The  the functions $\mathcal{T}_{\lambda_2\lambda_3\lambda_4}^{F}$  and $\mathcal{T}_{\lambda_2\lambda_3\lambda_4}^{E}$ for the non-vanishing helicity configuration for are shown in Table \ref{table20}.

\begin{table}[H]
\begin{center}
  \begin{tabular}{  || c | c | c || }
    \hline \hline
    $\lambda_2,\,\lambda_3,\,\lambda_4$ & $\mathcal{T}_{\lambda_1,\,\lambda_2,\,\lambda_3}^{F}$ & $\mathcal{T}_{\lambda_1,\,\lambda_2,\,\lambda_3}^{E}$  \\ \hline
   \hline\hline
     $-,-,-$ & $\left(\frac{2(s_{q_1r_1}^2-m_{\tilde{t}}^2\tilde{m}^2)\la q_13\ra[3r_4]}{\sqrt{3}s_{q_1r_1}\tilde{m}\la r_43\ra}\right)F_3$ & $\frac{2\la q_13\ra[3r_4]}{\sqrt{2}\la r_43\ra}F_3$\\ \hline
     $-,-,0$ & $\left(\frac{\sqrt{2}(s_{q_1r_1}^2-m_{\tilde{t}}^2\tilde{m}^2)\la q_1r_4\ra[3r_4]}{\sqrt{3}s_{q_1r_1}\tilde{m}^2}\right)F_3$ & $\frac{\la q_1r_4\ra[3r_4]}{\tilde{m}}F_3$ 
     \\ \hline
      $+,-,-$ & $\left(\frac{2(s_{q_1r_1}^2-m_{\tilde{t}}^2\tilde{m}^2)\la r_13\ra[3r_4]}{\sqrt{3}s_{q_1r_1}\tilde{m}\la r_1q_1\ra\la r_43\ra}\right)F_4$ &  $\frac{2\la r_13\ra[3r_4]}{\sqrt{2}\la r_1q_1\ra\la r_43\ra}F_4$
      \\ \hline
       $+,-,0$ & $\left(\frac{\sqrt{2}(s_{q_1r_1}^2-m_{\tilde{t}}^2\tilde{m}^2)\la r_1r_4\ra[3r_4]}{\sqrt{3}s_{q_1r_1}\tilde{m}^2\la r_1q_1\ra}\right)F_4$ & $\frac{\la r_1r_4\ra[3r_4]}{\tilde{m}\la r_1q_1\ra}F_4$
       \\ \hline
\hline
  \end{tabular}
      \caption{Expressions for the  functions $\mathcal{T}_{\lambda_2\lambda_3\lambda_4}^{F}$ and $\mathcal{T}_{\lambda_2\lambda_3\lambda_4}^{E}$ that appear in the amplitude of  the three-body stop decay $\tilde{t}(p_1)\to\tilde{\Psi}^{\mu}(p_2)b(p_3)W(p_4)$.}
  \label{table20}
 \end{center}
\end{table}

 \vspace{-0.18\linewidth}
\begin{minipage}{\linewidth}
\begin{figure}[H]
\centering
\begin{picture}(-60,75)
\put(-70,0){\includegraphics[scale=0.6]{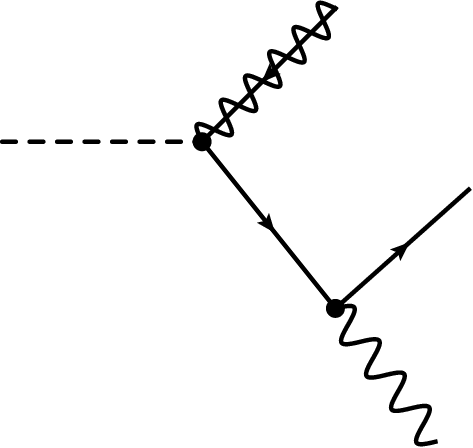}}
\put(-63,35){$\tilde{t}(p_1)$}
\put(-22,0){$\epsilon^{\nu}(p_4)$}
\put(-34, 45){$\tilde{\Psi}^{\mu}(p_2)$}
\put(-20, 25){$b(p_3)$}
\put(-42,24){$t(l)$}
\end{picture}
\caption{Feynman diagram for the three-body stop decay}
        \label{fig:tbsd}
\end{figure}
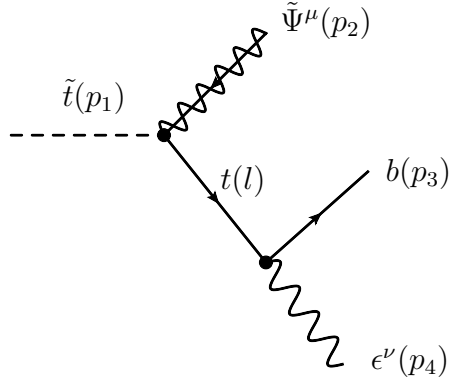
      \end{minipage}
     
       \vspace{0.03\linewidth}

 In Table \ref{table20}, we have use the following  definitions: 
 
\begin{align}
A_{\tilde{\Psi}}&=\frac{1}{s_{q_1r_1}}(s_{q_1r_1}-\tilde{m}^2),\\
A_{\tilde{t}}&=\frac{1}{s_{q_1r_1}}(s_{q_1r_1}-m_{\tilde{t}}^2),\\
F_3&=m_t\cos\theta_{\tilde{t}}-A_{\tilde{t}}\tilde{m}\sin\theta_{\tilde{t}},\\
F_4&=A_{\tilde{\Psi}}s_{q_1r_1}\sin\theta_{\tilde{t}}+m_t\tilde{m}\cos\theta_{\tilde{t}}.
\end{align}

  \subsection{Four-body stop decay $\tilde{t}\to\tilde{\Psi}^{\mu}\,b\,l\,\nu_{l}$}
  
In this case, we consider the Feynman diagram shown in Figure \ref{fig:s4bsd}.  The amplitude  for the four-body stop decay $\tilde{t}(p_1)\to\tilde{\Psi}^{\mu}(p_2)\,b(p_3)\,l(p_4)\,\nu_{l}(p_5)$ is written as follows

\begin{align}
\mathcal{M}_{\lambda_2\lambda_3\lambda_4\lambda_5}&=C_4P_t(l)P_{W}(q)\mathcal{T}_{\lambda_2\lambda_3\lambda_4\lambda_5},
\end{align}

the function $\mathcal{T}_{\lambda_2\lambda_3\lambda_4\lambda_5}^{F}$ for the full gravitino is given by

\begin{align}
\mathcal{T}_{\lambda_2\lambda_3\lambda_4\lambda_5}^{F}&=
\left(\overline{\tilde{\Psi}}_{\lambda_2}^{\beta}(p_2)\g_{\alpha}\g_{\beta}p_1^{\alpha}(\mathbf{P_R}\cos\theta_{\tilde{t}}+\mathbf{P_L}\sin\theta_{\tilde{t}})\right)\left(-\s{l}+m_t\right)(\g^{\nu}\mathbf{P_R}u_{\lambda_3}(p_3))\\
&\quad
\left(\eta_{\nu\mu}+\frac{q_{\mu}q_{\nu}}{M_W^2}\right)\left(\bar{u}_{\lambda_4}(p_4)\g^{\mu}\mathbf{P_L}v_{\lambda_5}(p_5)\right),
\end{align}

with $C_4^F=\frac{g_W^2}{\sqrt{3}M}$. We have defined  $P_W(q)=\frac{1}{q^2+M_W^2}$ as the denominator of the $W$ boson propagator (Figure \ref{fig:s4bsd}).

For the goldstino approximation the function $\mathcal{T}_{\lambda_2\lambda_3\lambda_4\lambda_5}^{E}$ takes the following form 

\begin{align}
\mathcal{T}_{\lambda_2\lambda_3\lambda_4\lambda_5}^{E}&=
\left(\bar{u}_{\lambda_2}(p_2)(\mathbf{P_R}\cos\theta_{\tilde{t}}+\mathbf{P_L}\sin\theta_{\tilde{t}})\right)\left(-\s{l}+m_t\right)(\g^{\nu}\mathbf{P_R}u_{\lambda_3}(p_3))\\
&\quad
\left(\eta_{\nu\mu}+\frac{q_{\mu}q_{\nu}}{M_W^2}\right)\left(\bar{u}_{\lambda_4}(p_4)\g^{\mu}\mathbf{P_L}v_{\lambda_5}(p_5)\right),
\end{align}
with $C_4^{E}=\frac{g_W^2(m_t^2-m_{\tilde{t}}^2)}{2\sqrt{3}M\tilde{m}}$.

 The functions $\mathcal{T}_{\lambda_2\lambda_3\lambda_4\lambda_5}^{F}$  and $\mathcal{T}_{\lambda_2\lambda_3\lambda_4\lambda_5}^{E}$ corresponding to the  non-vanishing helicity configuration are shown in Table \ref{table22}.

\begin{table}[H]
\begin{center}
  \begin{tabular}{  || c | c | c || }
    \hline \hline
    $\lambda_2,\,\lambda_3,\,\lambda_4,\,\lambda_5$ & $\mathcal{T}_{\lambda_2,\,\lambda_3,\,\lambda_4\,\lambda_5}^{F}$ & $\mathcal{T}_{\lambda_2,\,\lambda_3,\,\lambda_4\,\lambda_5}^{E}$
     \\ \hline
   \hline\hline
     $-,-,-,+$ & $\left(\frac{2\sqrt{2}(s_{q_1r_1}^2-m_{\tilde{t}}^2\tilde{m}^2)\la q_14\ra[35]}{\sqrt{3}s_{q_1r_1}\tilde{m}}\right)F_3$ & $2\la q_14\ra[35]F_3$
     \\ \hline
   $+,-,-,+$ & $\left(\frac{2\sqrt{2}(s_{q_1r_1}^2-m_{\tilde{t}}^2\tilde{m}^2)\la r_14\ra[35]}{\sqrt{3}s_{q_1r_1}\tilde{m}\la r_1q_1\ra}\right)F_4$ &  $\frac{2\la r_14\ra[35]}{\la r_1q_1\ra}F_4$
   \\ \hline
      \hline
    \hline
  \end{tabular}
      \caption{Expressions for the functions $\mathcal{T}_{\lambda_2\lambda_3\lambda_4\lambda_5}^{F}$ and $\mathcal{T}_{\lambda_2\lambda_3\lambda_4\lambda_5}^{E}$ that appear in the amplitude of the the four-body stop decay $\tilde{t}(p_1)\to\tilde{\Psi}^{\mu}(p_2)\,b(p_3)\,l(p_4)\,\nu_{l}(p_5)$.}
  \label{table22}
 \end{center}
\end{table}

\vspace{0.0093\linewidth}
\begin{minipage}{\linewidth}
\begin{figure}[H]
\centering
\begin{picture}(-60,75)
\put(-70,0){\includegraphics[scale=0.6]{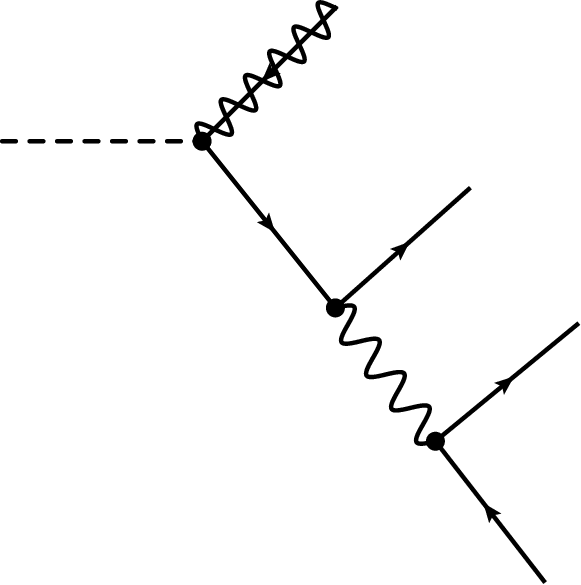}}
\put(-63,48){$\tilde{t}(p_1)$}
\put(-34, 57){$\tilde{\Psi}^{\mu}(p_2)$}
\put(-20, 40){$b(p_3)$}
\put(-9,26){$\nu(p_4)$}
\put(-12,0){$l^{+}(p_5)$}
\put(-42,38){$t(l)$}
\put(-30,24){$W(q)$}
\put(70,10){$\Psi(q)$}
\put(140,52){$V(\tilde{t}\Psi t)=\left(\frac{m_{t}^2-m_{\tilde{t}}^2}{\sqrt{3}Mm_{\tilde{G}}}\right)(\cos\theta\ \bold{P_R}+\sin\theta\ \bold{P_L})$}
\end{picture}
\caption{Feynman diagram for the four-body stop decay}
        \label{fig:s4bsd}
\end{figure}
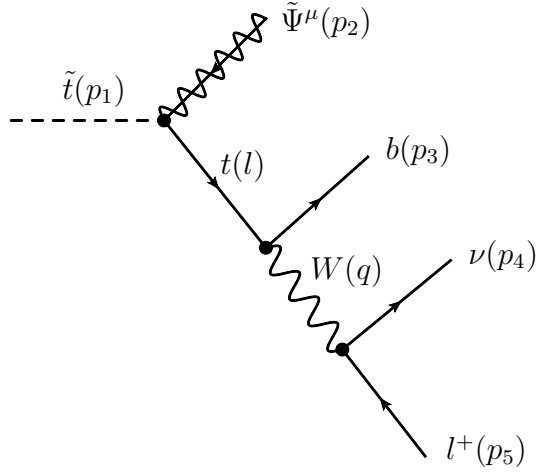
      \end{minipage}
     
       \vspace{0.03\linewidth}

\section{Numerical Results for stop decays}\label{numerical}
 \subsection{Two-body stop decay}
 In this section we shall present plots for the stop lifetime, in each of the cases considered. We shall also compare the gravitino and goldstino approximation. 
The decay width  for the process $\tilde{t}(p_1)\to \tilde{\Psi}^{\mu}(p_2)t(p_3)$ with massive gravitino is given as follows
\begin{align}
\Gamma_{\tilde{t}\to \tilde{G}\,t}
&=\frac{1}{48\pi M^2m_{\tilde{G}}^2m_{\tilde{t}}^3}\left((-m_{\tilde{G}}^2+m_t^2+m_{\tilde{t}}^2)^2-4m_t^2m_{\tilde{t}}^2\right)^{3/2}(m_{\tilde{t}}^2-m_{\tilde{G}}^2-m_t^2+2\sin2\theta\,m_tm_{\tilde{G}}).
\end{align}
On the other hand, when one employs the  goldstino approximation, the decay width  takes the form
\begin{align}
\Gamma_{\tilde{t}\to G\,t}
&=\frac{1}{48\pi M^2m_{\tilde{G}}^2m_{\tilde{t}}^3}\left((-m_{\tilde{G}}^2+m_t^2+m_{\tilde{t}}^2)^2-4m_t^2m_{\tilde{t}}^2\right)^{1/2}(m_{\tilde{t}}^2-m_t^2)^2\nonumber\\
&\quad\times(m_{\tilde{t}}^2-m_{\tilde{G}}^2-m_t^2+2\sin2\theta\,m_tm_{\tilde{G}}).
\end{align}

\begin{figure}[H]
\begin{minipage}[c]{0.42\linewidth}
\includegraphics[width=\linewidth]{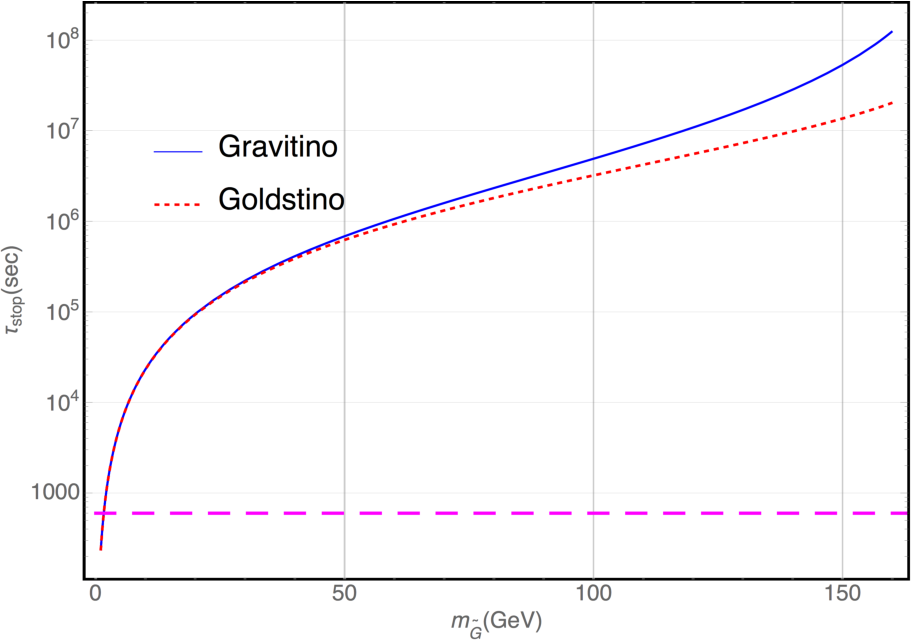}
\caption{The large dashed  horizontal plot represents the time (10 min) when the Big Bang nucleosynthesis has taken place.}
\label{2bd_stop_lt_fig1}
\end{minipage}
\hfill
\begin{minipage}[c]{0.43\linewidth}
\includegraphics[width=\linewidth]{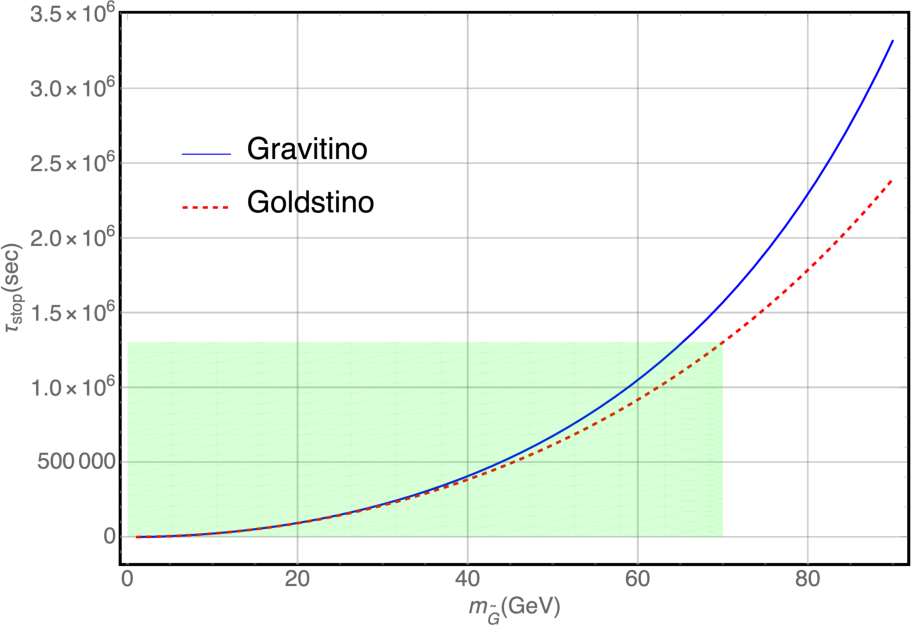}
\caption{In the shadow region the lifetime for the gravitino and goldstino approximation is smaller than $1\%$}
\label{2bd_stop_lt_fig2}
\end{minipage}%
\end{figure}

In Figure \ref{2bd_stop_lt_fig1} we have fixed the stop mass to $m_{\tilde{t}}=350\,GeV$ (the same for three- and four-body decay). Continuos line represent the stop lifetime with gravitino in the final state, while the dashed line represent the stop lifetime within the goldstino approximation. 
In the shadow region of Figure \ref{2bd_stop_lt_fig2}  (green for online version) the difference between the gravitino and  goldstino approximation is smaller than $1\%$. 

We compare the decay lifetimes  using the full gravitino with  the goldstino approximation. For masses bellow about $\tilde{m}<70$ GeV, the difference in lifetimes between the gravitino and goldstino approximations is less than $1\%$. 

Thus, we find that the stop squark is long-lived, and for the region of parameters where the 2-body mode
occurs, the lifetimes turns out to be  $\mathcal{O}$ ($10^8$) s. 
Thus, one can say that the 2-body mode is safe regarding considerations coming from 
big bang nucleosynthesis.

\subsection{Three- and Four-body stop decays}

\begin{figure}[H]
\begin{minipage}[c]{0.4\linewidth}
\includegraphics[width=\linewidth]{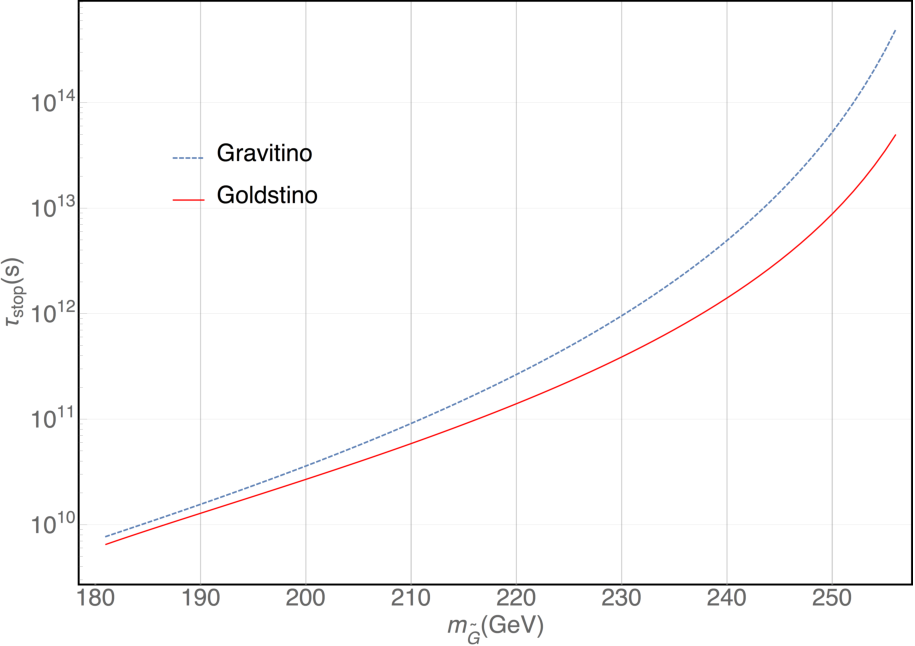}
\caption{Log plot for the lifetime of the three-body stop decay.}
\label{3bdp1}
\end{minipage}
\hfill
\begin{minipage}[c]{0.43\linewidth}
\includegraphics[width=\linewidth]{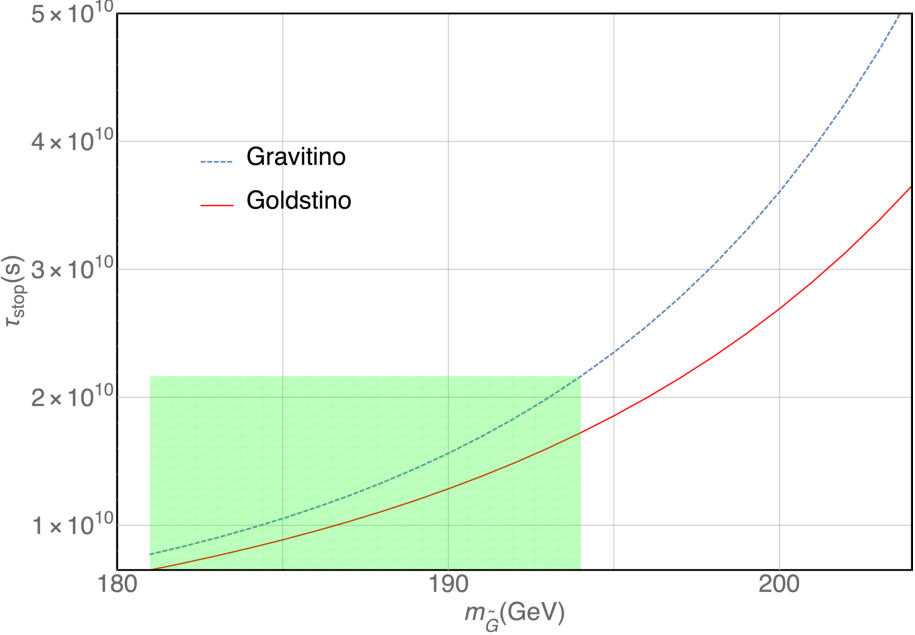}
\caption{Close up for the three-body stop decay.}
\label{3bdzp3}
\end{minipage}%
\end{figure}

In Figure (\ref{3bdp1}) we show the stop lifetime for the three-body mode using gravitino and goldstino approximation.  We notice that it takes values as large as $\mathcal{O}(10^{14}\,s)$. Figure (\ref{3bdzp3}) shows a close-up of the region where the difference in lifetimes between gravitino and goldstino approximation is smaller than $20\%$. This happens for gravitino masses lower than 194 GeV, for the chosen value of the stop mass ($m_{\tilde{t}}=350\,GeV$). 

\begin{figure}[H]
\begin{minipage}[c]{0.4\linewidth}
\includegraphics[width=\linewidth]{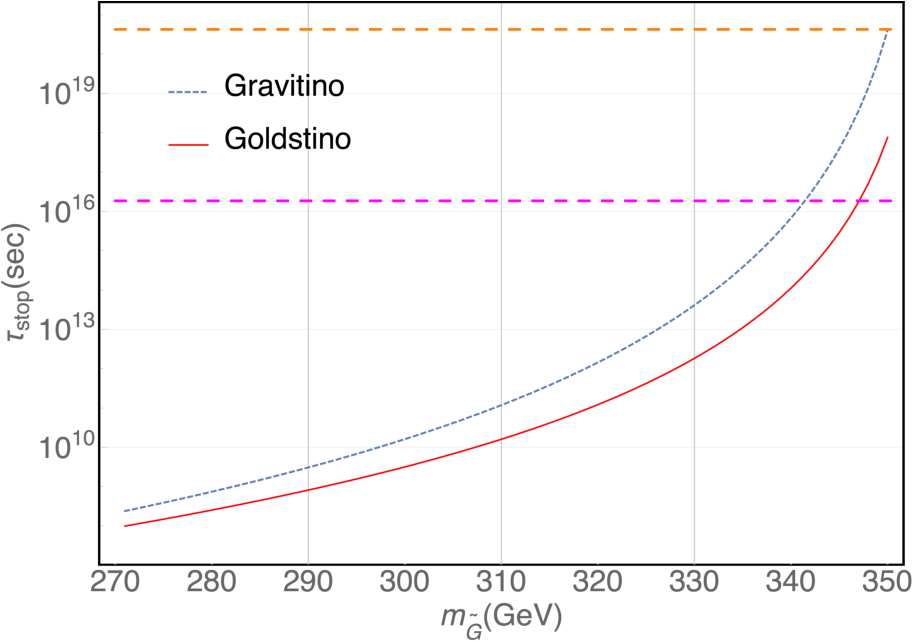}
\caption{Log plot for the lifetime of the four-body stop decay.}
\label{fbdp}
\end{minipage}
\hfill
\begin{minipage}[c]{0.43\linewidth}
\includegraphics[width=\linewidth]{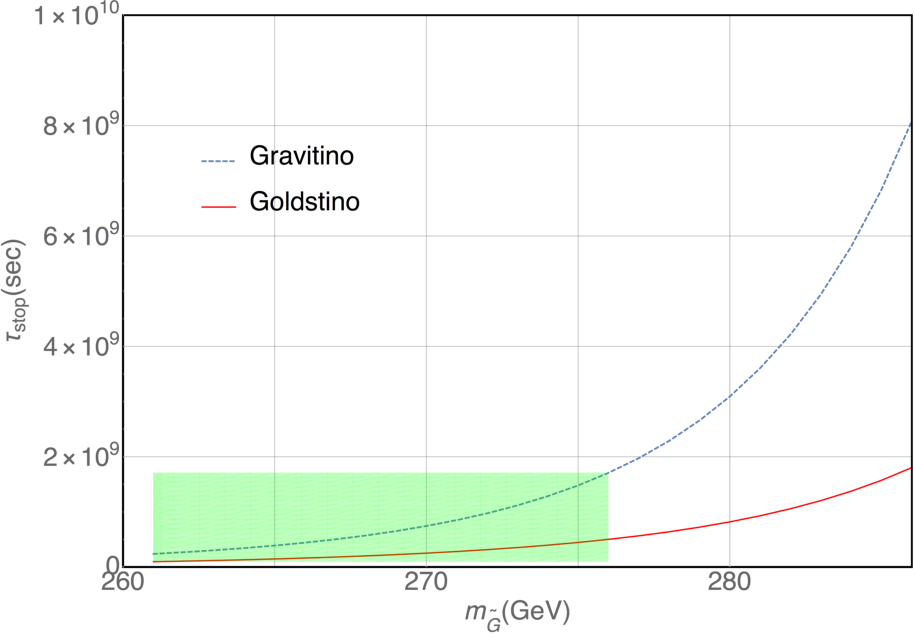}
\caption{Close up for the four-body stop decay.}
\label{fbdpcu}
\end{minipage}%
\end{figure}

Finally, we present in Figure (\ref{fbdp})  the result for the stop lifetime when  the four-body mode is the allowed channel. In this case, the stop lifetime can reach values up to  $\mathcal{O}(10^{20}\,s)$ which can have interesting consecuences. Figure (\ref{fbdpcu}) shows a close-up of the region where the difference in lifetimes between  gravitino and goldstino approximation is smaller than $70\%$. This happens for gravitino masses lower than 276 GeV, for the chosen value of the stop mass ($m_{\tilde{t}}=350\,GeV$).

Thus, both the 3- and 4-body mode, which reaches lifetimes of order $\mathcal{O}(10^{14}\,s)$ and $\mathcal{O}(10^{20}\,s)$. respectively, are safe regarding the constraints imposed by big bang nucleosynthesis.

However, in the 4-body mode, we find that it is possible that the decay  occurs during the epoch of galaxy formation (this is show as the magenta dashed line in Figure \ref{3bdzp3}, for completeness we also show the line representing the lifetime of the universe), which may have some impact from its decay products. 
Furthermore, as it can be seen from the figures, we also find  an small window of parameters where 
the stop lifetime could be even longer that the age of the universe, which will make the stop to be stable on cosmological scales. However, if it lives so long it would contribute to dark matter relic density.
But the stop is not dark at all, as it has both color and electric charges, and 
it will not fulfill the conditions to be dark matter, and therefore it appears to be 
excluded.

\section{Conclusions}\label{conclusions}

Thus paper contains an study of the stop ($\tilde{t}$) lifetime, taking into account the stop decay  into gravitinos ($\tilde{\Psi}_{\mu}$) in the final state, namely $\tilde{t}\to \tilde{\Psi}_{\mu} + X$, where $X=t, bW, b l\nu_l$ for the two-, three- and four-body decay modes, respectively. 
We have compared the full calculation obtained using the Feynman rules for massive gravitino, 
with the results obtained employing  the equivalence theorem, where the longitudinal component of the gravitino is replaced by the goldstino, an approximation that works very well for light gravitinos.

We find that the stop squark is very long-lived, with lifetimes of $\mathcal{O}$($10^8$s, $10^{14}$s, $10^{20}$s) for the 2-, 3- and 4-body modes under consideration. 
It is found that the lifetime obeys  the hierarchy: 
$\tau_{2b} ({\tilde t})  < \tau_{3b} ({\tilde t}) < \tau_{4b} ({\tilde t})$, as it should be. Thus, even the 2-body mode, which reaches lifetimes of order $10^8$s, is safe regarding big bang nucleosynthesis. However, in the 4-body mode
it is possible that the decay  occurs during the epoch of galaxy formation, which may have some impact. 
Furthermore, there is an small window of parameters where the stop lifetime could be longer that the age of
the universe, which will make the stop to be stable on cosmological scales, and being color and electrically
charged, it will not fulfill the conditions to be dark matter, and therefore it should be excluded.

At colliders, a long-lived  stop will have interesting signatures, in particular once it is produce it will hadronize and decay outside of the detector. This happens even for the lowest values of stop lifetime arising within of our scenarios.

\appendix
\section{Helicity Amplitudes}
\label{Basics}
In this appendix, we introduce the properties for the massless spinors that are used throughout this paper, most of them were taking from Ref.~\cite{srednicki}.  

Using the  spinor bra-ket notation, the 4-component Dirac spinor are rewritten as  follows 

\begin{align}\label{eq:bknotation}
u_-(p)&=v_{+}(p)=|p],\\
u_+(p)&=v_-(p)=|p\rangle,\\
\bar{u}_+(p)&=\bar{v}_-(p)=[p|,\\
\bar{u}_-(p)&=\bar{v}_+(p)=\langle p|,
\end{align}

which obey the following relations

\begin{align}
u_{s}(p)\bar{u}_s(p)&=\frac{1}{2}(1+s\g_5)(-\s{p}),\\
v_{s}(p)\bar{v}_s(p)&=\frac{1}{2}(1-s\g_5)(-\s{p}),
\end{align}

where $s=\pm$ indicates the helicity. The spinor products are antisymmetric, using the bra-ket notation this reads as follows

\begin{align}\label{eq:antys1}
	\bar{u}_+(p)u_-(k)&=[pk] =-[kp]=-\bar{u}_+(k)u_-(p), \\
	\bar{u}_-(p)u_+(k)&=\langle pk\rangle=-\langle kp\rangle=\bar{u}_-(k)u_+(p),\label{eq:antys2}
\end{align}

furthermore, taking the last results Eqs.~(\ref{eq:antys1})-(\ref{eq:antys2}) into account, one also have that the spinor product fulfill  $[qq]=\langle qq\rangle=0$, the type of spinor products $[k p \rangle$ and $\langle pk]$ are also null.  

For real momenta these spinor products satisfy

\begin{align}
	\langle pk\rangle&=[kp]^\ast,\\
	[kp]&=\langle pk\rangle^*,\\
[pq]\langle pq\rangle&=s_{pq}=-(p+q)^2=-2p\cdot q.
\end{align}

Another useful properties are the following

\begin{align}
	[k|\gamma^\mu|p\rangle&=\langle p|\gamma^\mu|k] , \label{sq_br_1} \\
	[k|\gamma^\mu|p\rangle^\ast&=[p|\gamma^\mu|k\rangle,  \label{sq_br_2} \\
	\langle p|\slashed{k}|q]&=-\langle pk\rangle[kq],\\
	 	\langle p|\gamma^\mu|p]&=2p^\mu.
	 \end{align}

The Fierz identity is also a useful property, this take the following form 

\begin{equation}
	\langle p|\gamma^\mu|q]\langle r|\gamma_\mu|w]=2\langle pr\rangle[qw] .
\end{equation}

From the completeness relation, one is able to express $\s{p}$ as a  product of spinors, this is as follows

\begin{equation}
\s{p}=-(|p]\langle p|+|p\rangle[p|).
\end{equation}

\section*{Acknowledgments}
We would like to acknowledge the support of CONACYT and SNI. BL acknowledge the support from UNAH and MCTP as well as the hospitality of ICTP where part of this work has been done.

\end{document}